\documentclass[aps,prd,twocolumn,eqsecnum,showpacs,amsmath]{revtex4}
\usepackage{graphicx}
\begin{document}

\thispagestyle{empty}

\title{Kink instability and stabilization of the Friedmann universe
with scalar fields}
\author{
Hideki Maeda $^{1}$\footnote{Electronic
 address:hideki@gravity.phys.waseda.ac.jp} and
Tomohiro Harada $^{2}$\footnote{Electronic
 address:T.Harada@qmul.ac.uk}}
\affiliation{
$^{1}$Advanced Research Institute for Science and Engineering,
Waseda University, Okubo 3-4-1, Shinjuku, Tokyo, 169-8555, Japan\\
$^{2}$Astronomy Unit, School of Mathematical Sciences,
Queen Mary, University of London,
Mile End Road, London E1 4NS, UK
}
\date{\today}

\begin{abstract}                
The evolution of weak discontinuity is investigated in
the flat FRW universe with a single scalar field and with multiple scalar 
fields. We consider both massless scalar fields and scalar fields with 
exponential potentials. Then we find that a new type of instability, 
i.e. kink instability develops in the flat FRW universe with massless 
scalar fields. The kink instability develops with scalar 
fields with considerably steep exponential potentials, while 
less steep exponential potentials do not suffer from kink instability. 
In particular, assisted inflation with multiple scalar fields
does not suffer from kink instability. 
The stability of general spherically symmetric self-similar solutions 
is also discussed.
\end{abstract}

\maketitle

\section{introduction}
Future precision observations of the cosmic microwave background radiation and/or gravitational lensing phenomena will provide a large number of new data to be analyzed and interpreted. 
The standard cosmology assumes that the observable part of the universe on scales of the order of the present Hubble distance is almost spatially homogeneous and isotropic. Actually, the data of WMAP for the cosmic microwave background radiation tell us that the universe at the last scattering surface is well approximated by the Friedmann-Robertson-Walker (FRW) universe so that understanding of the detailed properties of the FRW universe is an important step in cosmology~\cite{wmap}. 

One of the most important problems is how cosmological structures on various scales formed from the very homogeneous and isotropic universe at early times. It is known that the gravitational instability scenario with cold dark matter 
well describes structure formation from small primordial perturbations in our universe~\cite{mfb1992}. 

Recently, a new kind of instability has been
found in the studies of self-similar solutions. Self-similar solutions
are defined 
in terms of a homothetic Killing vector field in general
relativity~\cite{ct1971} and include a class of the flat FRW
solutions. The instability concerns weak discontinuity. If the
perturbations with weak discontinuity are inserted into a self-similar
solution which is unstable against this mode, the discontinuity grows as
time proceeds. This instability has been called {\em kink
instability}. Kink instability of self-similar solutions 
was originally found by Ori and Piran in
spherically symmetric isothermal gas systems in Newtonian
gravity~\cite{op1988b}. In general relativity, Harada investigated the
kink instability of self-similar solutions for the spherical system of a
perfect fluid with the equation of state $p=k\mu$~\cite{harada2001}. The present authors investigated the kink
instability of self-similar solutions for the spherical system of a
stiff ($k=1$) fluid and those of a massless scalar
field~\cite{hm2003}. These works have shown that the kink instability
can occur in a large class of self-similar solutions. In particular, the
work in general relativity showed that the flat FRW universe is
unstable for a perfect fluid for $1/3 \le k \le 1$ and 
for a massless scalar field. 

In this letter we derive a stability criterion against kink mode
perturbations for the power-law flat FRW universe with a 
single scalar field and with multiple scalar fields with and without 
exponential potentials, 
which arise naturally in supergravity~\cite{townsend2001} 
or theories obtained through 
dimensional reduction to effective four dimensional
theories~\cite{gl2000,eg2003}. The stability of general spherically symmetric self-similar solutions with scalar fields is also discussed in Appendix~\ref{general}. 
We adopt the units such that $c=G=1$
and the abstract index notation of~\cite{wald1983}.

\section{stability criterion for the flat FRW universe}
\label{body}
\subsection{power-law flat FRW solution with and without exponential 
potentials}
\label{FRWbasic}
We begin with the action which describes 
the self-gravitating system of scalar fields with and without 
exponential potentials. 
\begin{equation}
S=\int d^4x\sqrt{-g}\left\{\frac{1}{8\pi}R-\sum^n_{i=1}\left[\frac12 (\nabla \phi_{i})^2+V_{i}e^{-\sqrt{8\pi} \lambda_{i}\phi_{i}}\right]\right\},
\label{action}
\end{equation}
where $\{V_i|i=1,\cdots,n\}$ and 
$\{\lambda_{i}|i=1,\cdots,n\}$ 
are $n$ real non-negative and positive constants, respectively. 
In this model we currently have $n$ real scalar fields 
$\{\phi_{j}|j=1,\cdots,n\}$, for which the total potential has the form 
\begin{eqnarray}
V_{\rm tot} \equiv \sum^n_{i=1}V_{i}e^{-\sqrt{8\pi} \lambda_{i}\phi_{i}}.
\end{eqnarray}
If $V_{j}=0$, then $\phi_{j}$ is massless and $\lambda_{j}$
is meaningless. 
We assume $0 < \lambda_1 \le \lambda_2 \le \cdots \le \lambda_n$ without loss of generality. This action has been considered in the context of assisted inflation~\cite{lms1998,mw1999,cvdh2000}. The action involving a perfect fluid has been studied by several authors~\cite{cvdh2000,gpcz2003}. 

Then the total energy-momentum tensor is given by 
\begin{equation}
T^{ab}=\sum^n_{i=1}\left(\nabla^{a}\phi_{i}\nabla^{b}\phi_{i}-\frac{1}{2}g^{ab}\nabla_{c}\phi_{i}\nabla^{c}\phi_{i}\right)-g^{ab}V_{\rm tot},
\label{eq:stress-energy_tensor_of_scalar_field}
\end{equation}
while equations of motion for $\{\phi_j\}$ is given by 
\begin{eqnarray}
\nabla^{a}\nabla_{a} \phi_{j}=\frac{\partial V_{\rm tot}}{\partial \phi_{j}}.\label{eq:Klein-Gordon1}
\end{eqnarray}

We consider the power-law flat FRW spacetime with the metric;
\begin{equation}
ds^2=-dt^{2}+\left(\frac{t}{t_0}\right)^{2\alpha}(dr^2+r^{2}d\Omega^2),\label{flatFRW1}
\end{equation}
where $\alpha$ and $t_0$ are constants and $d\Omega^2 \equiv d\theta^{2}+\sin^{2}\theta d\varphi^2$. 

We first show that there are no power-law
flat FRW solutions which include both nontrivial 
massless scalar fields and scalar fields with exponential potentials.
When there exist $m$ scalar fields
$\{\phi_j|1 \le j\le m\}$ with a total potential 
$\sum^m_{i=1}V_i \exp(-\sqrt{8\pi}\lambda_i \phi_i)$ and $(n-m)$ massless scalar fields $\{\phi_j|m+1 \le j\le n\}$, the Einstein equations $G_{ab}=8\pi T_{ab}$ and the equations of motion for scalar fields are written as
\begin{eqnarray}
-3\alpha(\alpha-1)&=&8\pi t^2\biggl[\sum^n_{i=1}\phi_{i,t}^2 \nonumber \\
&&-\sum^m_{i=1}V_i e^{-\sqrt{8\pi}\lambda_i \phi_i}\biggl],\label{feq1}\\
\alpha(3\alpha-1)&=&8\pi t^2\sum^m_{i=1}V_i e^{-\sqrt{8\pi}\lambda_i \phi_i},\label{feq2}\\
t^2\phi_{j,tt}+3\alpha t\phi_{j,t}&=&\left\{
\begin{array}{ll}
\sqrt{8\pi}\lambda_j V_j t^2 e^{-\sqrt{8\pi}\lambda_j \phi_j}\\ \label{feq3}
~~~~~~~~~~~~(\mbox{for}~1 \le j\le m),\\ 
0~~~~~~(\mbox{for}~m+1 \le j\le n),
\end{array}
\right.
\end{eqnarray}
where the comma denotes the partial derivative. 
From Eqs.~(\ref{feq1}) and (\ref{feq2}), we obtain
\begin{eqnarray}
\alpha=4\pi t^2\sum^n_{i=1}\phi_{i,t}^2,\label{feq4}
\end{eqnarray}
so that $\{\phi_j\}$ must have the form 
\begin{eqnarray}
\phi_j=-\kappa_j \ln t +C_j,
\end{eqnarray}
where $\{\kappa_j\}$ are constants satisfying
\begin{eqnarray}
\alpha=4\pi \sum^n_{i=1}\kappa_{i}^2,\label{feq5}
\end{eqnarray}
and $\{C_j\}$ are constants. 
We can set $C_j=0$ for all $j$ by redefining the scalar field.
From Eq.~(\ref{feq2}) and the assumption of $V_j>0$ for $1 \le j \le m$,
$\kappa_j=-2/(\sqrt{8\pi}\lambda_j)$ must be satisfied for 
$1 \le j \le m$. 
Then Eqs.~(\ref{feq2}) and (\ref{feq3}) reduce to 
\begin{eqnarray}
\alpha=\sum^m_{i=1}\frac{2}{\lambda_i^2}.\label{feq6}
\end{eqnarray}
On the other hand, we obtain from Eq.~(\ref{feq5}) that 
\begin{eqnarray}
\alpha=\sum^m_{i=1}\frac{2}{\lambda_i^2}+4\pi \sum^n_{i=m+1}\kappa_i^2.\label{feq7}
\end{eqnarray}
Eqs.~(\ref{feq6}) and (\ref{feq7}) give a contradiction $\kappa_j=0$ for
$m+1\le j \le n$. Therefore, we conclude that 
if the power-law flat FRW solution includes both 
massless scalar fields and scalar fields with exponential potentials,
those massless scalar fields must be constant, i.e., trivial.

When there exist only massless scalar fields, the Einstein equations and the equations of motion for scalar fields give $\alpha=1/3$ and $4\pi \sum^n_{i=1}\kappa_i^2=1/3$, while when there are no massless scalar fields, they give 
\begin{eqnarray}
\alpha&=&\sum^n_{i=1}\frac{2}{\lambda_i^2},\\
V_j&=&\frac{1}{4\pi \lambda_j^2}\left(\sum^n_{i=1}\frac{6}{\lambda_i^2}-1\right).
\end{eqnarray}
The latter equation gives a condition $\sum^n_{i=1}\lambda_i^{-2}>1/6$ 
for $V_j>0$. Setting $n=1$ in both cases, we can obtain the results for a single scalar-field case.

For the later convenience, we transform the metric (\ref{flatFRW1}) to the self-similar form~(\ref{ssform}). As shown in
Appendix~\ref{ssFRW}, 
the flat FRW solution is self-similar if and only if the scale factor 
obeys a power-law.
Therefore it is concluded that there are no self-similar flat FRW solutions 
if there exist both nontrivial massless scalar fields and 
those with exponential potentials. 
Because the case with $\alpha=1$ is exceptional, in which the homothetic Killing vector is not tilted~\cite{mh2004}, we consider only the case with $\alpha \ne 1$. With a coordinate transformation $r=|1-\alpha|^{-1}t_0^{\alpha}{\bar r}^{1-\alpha}$, the metric (\ref{flatFRW1}) is transformed to the self-similar form~(\ref{ssform}) as
\begin{equation}
ds^2=-dt^{2}+z^{2\alpha}(d{\bar r}^2+(1-\alpha)^{-2}{\bar r}^{2}d\Omega^2), \label{FRWss}
\end{equation}
where $z \equiv t/{\bar r}$ is the self-similarity coordinate. 

The particle horizon $z=1$ plays an important role 
in the following analysis. This is because it is 
a {\it similarity horizon}, which is given by a radial 
null ray on which $z$ is constant.

\subsection{kink instability}
We consider general full-order spherically symmetric perturbations 
with scalar fields such as 
\begin{eqnarray}
ds^2&=&-e^{2{\bar \sigma(t,{\bar r})}}dt^{2}+z^{2\alpha}e^{2{\bar \omega(t,{\bar r})}}d{\bar r}^2 \nonumber \\
&&+(1-\alpha)^{-2}z^{2\alpha}{\bar r}^{2}e^{2{\bar \eta(t,{\bar r})}}d\Omega^2,\label{perturb1}\\
\phi_j&=&-\kappa_j\ln t+\bar{\psi}_j(t,{\bar r}).\label{perturb2}
\end{eqnarray} 
Hereafter we adopt the isotropic coordinates, in which ${\bar \omega}={\bar \eta}$. We define the following coordinates:
\begin{eqnarray}
Z &\equiv& \ln z, \\
T &\equiv& \ln t,
\end{eqnarray}
in which Eqs.~(\ref{perturb1}) and (\ref{perturb2}) are represented by 
\begin{eqnarray}
ds^2&=&-(e^{2\sigma(T,Z)}-e^{2(\alpha-1)Z+2\omega(T,Z)})e^{2T}dT^2 \nonumber \\
&&-2e^{2(\alpha-1)Z+2T+2\omega}dTdZ+e^{2(\alpha-1)Z+2T+2\omega}dZ^2 \nonumber \\
&&+(1-\alpha)^{-2}e^{2(\alpha-1)Z+2T+2\omega}d\Omega^2,\\
\phi_j&=&-\kappa_jT+\psi_j(T,Z),
\end{eqnarray} 
where 
\begin{eqnarray*}
\sigma(T,Z) &\equiv& {\bar \sigma(t(T),{\bar r}(T,Z))}, \\
\omega(T,Z) &\equiv& {\bar \omega(t(T),{\bar r}(T,Z))}, \\
\psi_j(T,Z) &\equiv& {\bar \psi}_j(t(T),{\bar r}(T,Z)),
\end{eqnarray*}
for $1\le j\le n$.

We consider perturbations $\sigma$, $\omega$ and $\{\psi_j\}$ which satisfy the following conditions:\\
(1) The initial perturbations vanish outside the similarity horizon.  \\
(2) $\sigma$, $\omega$, $\{\psi_j\}$ and $\{\psi'_j\}$ are continuous, in particular at the similarity horizon $Z=0$, where a prime denotes the derivative with respect to $Z$. \\
(3) $\{{\dot \psi}''_j\}$ and $\{\psi''_j\}$ are discontinuous at the similarity horizon, although they have finite one-sided limit values as $Z\to -0$ and $Z\to +0$, where a dot denotes the derivative with respect to $T$.\\
(4) The quasi-local mass is continuous, i.e., there are no singular hypersurfaces in the spacetime. 

Now we consider the behavior of the perturbations at the similarity
horizon.
The perturbations satisfy $\psi_j=\psi'_j=0$ for
all $j$ and $\psi''_j\ne 0$ for some $j$ 
at the similarity horizon at the initial moment $T=T_{0}$ due to condition
(2). The evolution of
the initially unperturbed region is completely described by the
background flat FRW solution because information from the perturbed
side cannot penetrate the unperturbed side due to condition (1). Then, we
find $\sigma=\omega=0$, $\psi_j=\psi'_j=0$ for all $j$
and $\psi''_j \ne 0$ for some $j$ at the similarity horizon for $T\ge
T_{0}$ due to conditions (2) and (3). The Misner-Sharp quasi-local mass $m$ is defined by
\begin{eqnarray}
m=\frac{R}{2}(1-R_{,\mu}R^{,\mu}),
\end{eqnarray} 
where $R$ is the circumferential radius. Because of
\[
 R=|1-\alpha|^{-1}\exp{((\alpha-1)Z+T+\omega)}
\] 
in the present case, $\omega'=0$ is satisfied at the similarity horizon
due to condition (4). Then we find $\sigma'=0$ from the equation of
motion for scalar fields. From the ($00$)+($01$) and ($11$)+($01$)
components of the Einstein equations 
\[
 R_{ab}=8\pi\left(T_{ab}-\frac{1}{2}g_{ab}T\right)
\], $\omega''=0$ and $\sigma''=0$ are obtained, respectively. Differentiating the equation of motion for scalar fields with respect to $Z$ and estimating both sides at the similarity horizon, we obtain 
\begin{eqnarray}
\dot{\psi}_{j}^{''}-\left(1-2\alpha\right)\psi_{j}^{''}=\frac12\lim_{Z\to 0}\left(1-e^{2(\alpha-1)Z}\right)\psi_{j}^{'''},
\end{eqnarray}
for each $j$.
We can show that the right-hand side vanishes~\cite{hm2003}. Finally, the full-order perturbation equation for $\psi_{j}^{''}$ at the similarity horizon is obtained as
\begin{equation}
\dot{\psi}_{j}^{''}=\left(1-2\alpha\right)\psi_{j}^{''}.
\label{eq:kink_mode_equation_scalar_field}
\end{equation}
It should be noted that the perturbations are those of full-order although this equation is linear. This equation can be integrated to obtain
\begin{eqnarray}
\psi_{j}^{''} \propto e^{(1-2\alpha)T}.\label{eq:kink_mode_solution_for_scalar_field2}
\end{eqnarray}
Therefore, it is found that the perturbation decays exponentially for $\alpha >1/2$, it is constant for $\alpha=1/2$ and it grows exponentially for $\alpha <1/2$. 

It is noted that these perturbations are gauge-independent as shown in Appendix~\ref{gauge}. Here we define instability by the exponential growth of discontinuity. Then we find the following criterion: the flat FRW universe with $\alpha >1/2$ are stable against the kink mode, while those with $\alpha <1/2$ are unstable. Solutions with $\alpha=1/2$ are marginally stable against this mode. 

Although we concentrate on the flat FRW solutions in this letter, we can obtain the stability criterion for general spherically symmetric self-similar solutions with regular similarity horizons adopting the similar analysis as that in~\cite{hm2003}, which is summarized in Appendix~\ref{general}.


\section{Discussions}
We have derived the stability criterion for spherically symmetric self-similar solutions 
with scalar fields with independent exponential potentials to suffer
from kink instability. 
It can be applied to any spherically symmetric self-similar solutions with regular similarity horizons. 
The kink instability, which we have considered
here, was studied in general relativity for a perfect fluid with an
equation of state $p=k\mu$~\cite{harada2001,hm2003} and for a
massless scalar field~\cite{hm2003}. 
We focus on the flat FRW universe in this letter, for which result is summarized in Tables~\ref{tb:perfect_fluid}
and \ref{tb:scalar_field}. 
This is a new kind of instability of the FRW universe. 
We first discuss the relations of our result to other studies in the literature on the system with scalar fields with or without exponential potentials. 
After that, we discuss astrophysical or cosmological implications of the kink instability of the flat FRW universe.

Let us see the flat FRW universe filled with a single matter field. 
The FRW universe is unstable for a
perfect fluid with $1/3 \le k \le 1$, for a massless scalar field and
for a scalar field with an exponential potential for
$4<\lambda^2<6$. These results suggest that if there exists a phase in
the flat FRW universe in which a scalar field dominates other matter
fields, the scalar field with an exponential potential for
$0<\lambda^2<4$ is preferred rather than a massless scalar field. Such a
potential naturally arises in supergravity~\cite{townsend2001} or
theories obtained through
dimensional reduction to effective four 
dimensional theories~\cite{gl2000,eg2003}.

Kitada and Maeda~\cite{km1992,km1993} showed the cosmic no-hair theorem,
which is a generalization of the result by Wald~\cite{wald1983b} and
states that all ever-expanding homogeneous models which contain a scalar
field with an exponential potential together with a matter field
satisfying the dominant energy condition asymptote to the inflationary
flat FRW solution for $0<\lambda^2<2$. 
According to the criterion obtained in this letter, the flat FRW solution with $4<\lambda^2<6$ suffers from kink mode perturbations, which are kinds of inhomogeneous perturbations, so that our result does not affect the cosmic no-hair theorem.
 
Next let us move on to the case where the flat FRW universe is filled
with multiple scalar fields.
In such a situation, Liddle, Mazumdar and Schunck proposed assisted
inflation, where an arbitrary number of scalar fields with independent
exponential potentials evolve to the inflationary flat FRW solution even
if each individual potential is too steep to support inflation by its
own~\cite{lms1998}. Coley and van den Hoogen showed that the assisted
inflationary solution is a global attractor in the spatially homogeneous
and isotropic universe if
$\sum^n_{i=1}\lambda_i^{-2}>1/2$~\cite{cvdh2000}. Together with the
result on the spatially homogeneous scalar-field cosmological models by
Billyard {\it et al.}~\cite{bcvdhio1999}, they concluded that the
assisted inflationary solution is a global attractor for all
ever-expanding spatially homogeneous cosmological models with 
multiple scalar fields with exponential potentials provided $\sum^n_{i=1}\lambda_i^{-2}>1/2$~\cite{cvdh2000}. According to the criterion obtained in this letter, the expanding flat FRW solution with $\sum^n_{i=1}\lambda_{i}^{-2} >1/4$ does not suffer from kink instability, so that our result does not affect the genericity of assisted inflation even if we allow initial data with weak discontinuity in the second-order derivative of the scalar field. 

For any case of the flat FRW universes 
we have seen in this letter,
we find that the stability of the 
flat FRW universe against kink mode perturbation
changes from stability to instability as
the power index $\alpha$ of the scale factor increases
through $1/2$, which is clear from Tables~\ref{tb:perfect_fluid}
and \ref{tb:scalar_field}.
This suggests that the cosmological kink instability is related to 
the expansion-law of the universe rather than the characteristics
of the matter fields. 

The kink instability is particularly important 
for the flat FRW universe in the radiation ($k=1/3$) dominated era.
It is critical for some cosmological models 
which assume a stage when the effective equation of state
is ``harder'' than the radiation fluid. 
Because the flat FRW solution is homogeneous, 
we can choose any point as a symmetric centre 
in the present analysis.
This implies that a bubble-like structure will develop and the 
bubble walls collide with each other.
This consideration suggests that density perturbations can arise
in different scales as a result of multiple bubble wall collisions.
Although the product of kink instability has not been clear yet,
it is plausible that it will bring shock-wave formation, which could be important for the cosmological structure formation.
Actually, the shock-wave formation has been observed in the numerical simulations of the gravitational collapse of the density fluctuations in the radiation-dominated flat FRW universe~\cite{hs2002}. 
The outcome of the instability and the effect on the structure formation should be comprehensively investigated.


\acknowledgments
HM is grateful to T.~Hanawa and S.~Inutsuka for helpful discussions. TH is grateful to A.~Ishibashi for helpful discussions. This work was partially supported by a Grant for The 21st Century COE Program (Holistic Research and Education Center for Physics Self-Organization Systems) at Waseda University. TH was supported by JSPS Postdoctoral Fellowship for Research Abroad.

\appendix


\section{stability criterion for general self-similar solutions}
\label{general}
Here we show that kink instability may occur in a very large 
class of spherically symmetric self-similar spacetimes.
The analysis for the case of a single massless scalar
field~\cite{hm2003} is straightforwardly extended to the case of 
multiple scalar fields.

We adopt the Bondi coordinates for spherically symmetric spacetimes as
\begin{equation}
ds^{2}=-g\bar{g}du^{2}-2gdudR+R^{2}d\Omega^2,
\label{eq:Bondi_coordinates}
\end{equation}
where $g=g(u,R)$ and $\bar{g}=\bar{g}(u,R)$. 
For later convenience we define new functions $\{{\bar h_j(u,R)}\}$ as
\begin{eqnarray}
\phi_{j}(u,R)={\bar h_{j}(u,R)} -\kappa_{j}\ln|u|,
\end{eqnarray}
where $\{\kappa_{j}|j=1,\cdots,n\}$ 
are constants. We define the following self-similar coordinates:
\begin{equation}
x \equiv  -\frac{R}{u}, \quad X \equiv \ln|x|, \quad U \equiv  -\ln |u|.
\end{equation}
We refer to $u<0$ and $u>0$ as early time and late time, respectively. As $u$ is increased, $U$ increases in early times, while $U$ decreases
in late times. As $u$ is fixed and $R$ is increased, $X$ increases. 

Then the Einstein equation and the equations of motion for scalar fields reduce to the following partial differential equations:
\begin{eqnarray}
(\ln g)'&=&4\pi \sum^n_{i=1} {\bar h}_{i}^{'2},\label{basic1}\\
{\bar g}'&=&g\left(1-8\pi r^2\sum^n_{i=1}|u|^{\sqrt{8\pi}\lambda_i\kappa_i}V_{i}e^{-\sqrt{8\pi} \lambda_{i}{\bar h}_{i}}\right)\nonumber \\
&&-{\bar g},\label{basic2}\\
g\left(\frac{{\bar g}}{g}
\right)'
&=&-g\left(\frac{{\bar g}}{g}\right)^{\dot{}}+8\pi \sum^n_{i=1} \left({\dot {\bar h}}_{i}+{\bar h}'_{i}+\kappa_{i}\right)\nonumber \\
&& \times\left\{x\left({\dot {\bar h}}_{i}+{\bar h}'_{i}+\kappa_{i}\right)-{\bar g}{\bar h}'_{i}\right\},\label{basic3}\\
({\bar g}{\bar h}'_{j})'+{\bar g}{\bar h}'_{j}&=&-\sqrt{8\pi}\lambda_{j} gr^2|u|^{\sqrt{8\pi}\lambda_j\kappa_j} V_{j}e^{-\sqrt{8\pi} \lambda_{j}{\bar h}_{j}}\nonumber \\
&&+2x\left[({\dot {\bar h}}_{j}+{\bar h}'_{j})+({\dot {\bar h}}_{j}+{\bar h}'_{j})'+\kappa_{j}\right],\label{basic4}
\end{eqnarray}
where the dot and prime denote the partial derivatives with respect to $U$ and $X$, respectively.

\subsection{self-similar multi-scalar-field solutions}
\label{subsec:self-similar_scalar-field_solutions}
For self-similar solutions, we assume that $g=g(x)$,
${\bar g}={\bar g}(x)$ and ${\bar h}_{j}={\bar h}_{j}(x)$ for all $j$. 
From Eq.~(\ref{basic4}), $\kappa_{j}$ is arbitrary if $\phi_{j}$ is massless~\cite{brady1995}, while $\kappa_j=-2/(\sqrt{8\pi}\lambda_{j})$ should be satisfied otherwise for consistency~\cite{we1997}. 
Dropping the terms with dots in Eqs.~(\ref{basic1})--(\ref{basic4}), we obtain a set of ordinary differential equations for $g$, $\bar{g}$ and $\{\bar{h}_{j}\}$.

These equations are singular at $\bar{g}=2x$, which is called a {\it similarity horizon}. We denote the value of $x$ at the similarity horizon as $x_{\rm (s)}$ and also the value of $X$ as $X_{(s)}\equiv \ln |x_{(s)}|$. We consider self-similar solutions with finite values of functions $g$, $\bar{g}$, $\{\bar{h}_{j}\}$ and $\{\bar{h}_{j}'\}$ and their gradients with respect to $X$ at the similarity horizon. Then we find at the similarity horizon $X=X_{\rm (s)}$
\begin{eqnarray}
g_{\rm (s)} &=& \frac{x_{\rm (s)}(2+8\pi\sum^n_{i=1}\kappa_{i}^2)}{1-8\pi x_{\rm (s)}^2\sum^n_{i=1}V_{i}e^{-\sqrt{8\pi} \lambda_{i}{\bar h}_{i \rm (s)}}},\label{eq:regularity_similarity_horizon1} \\
{\bar g}_{\rm (s)}&=&2x_{\rm (s)},\label{eq:regularity_similarity_horizon2} \\
{\bar h}_{j{\rm (s)}}' &=& \frac{2\kappa_{j}-\sqrt{8\pi} \lambda_{j}x_{\rm (s)}g_{\rm (s)} V_{j}e^{-\sqrt{8\pi} \lambda_{j}{\bar h}_{j \rm (s)}}}{8\pi\sum^n_{i=1}\kappa_{i}^2}, \label{eq:regularity_similarity_horizon3}
\end{eqnarray}
for $\kappa_j \ne 0$, where the subscript ${\rm (s)}$ denotes the value at the similarity horizon. When $\kappa_j=0$ for all $j$, which can be satisfied in the case that all scalar fields are massless, i.e. $V_j=0$ for all $j$, we find 
\begin{eqnarray}
g_{\rm (s)}={\bar g}_{\rm (s)}=2x_{\rm (s)},~~{\bar h}_{j{\rm (s)}}' = {\bar h}_{j{\rm (s)}}', \label{eq:regularity_similarity_horizon22}
\end{eqnarray}
at the similarity horizon. A similarity horizon corresponds to a radial null curve because
\begin{equation}
\frac{dX}{dU}=1-\frac{\bar{g}}{2x} \label{innull}
\end{equation}
is satisfied along a radial null curve. No information propagate
inwardly beyond the similarity horizon in early-time (late-time)
solutions for $g_{\rm (s)}{\bar g}_{\rm (s)}>(<)0$, while no information
propagate outwardly beyond the similarity horizon in late-time
(early-time) solutions for $g_{\rm (s)}{\bar g}_{\rm (s)}>(<)0$. 

A class of the expanding flat FRW solutions belongs to the late-time
self-similar solutions with an analytic similarity horizon. 
The flat FRW solutions are characterized by the parameters $\{\kappa_j\}$, which are gauge-independent. 
Actually, the FRW solution in the same
coordinates with those used in this appendix was obtained by
Christodoulou for the case of a massless scalar 
field~\cite{christodoulou1994}. 


\subsection{kink instability of self-similar solutions}
\label{subsec:kink_scalar_field}
We consider perturbations which satisfy the following conditions in the background self-similar solution:\\
(1) The initial perturbations vanish inside the similarity horizon for early-time (late-time) solutions with $g_{\rm (s)}{\bar g}_{\rm (s)}>(<)0$. (Conversely, the initial perturbations vanish outside the similarity horizon for late-time (early-time) solutions with $g_{\rm (s)}{\bar g}_{\rm (s)}>(<)0$.) \\
(2) $g$, ${\bar g}$, $\{{\bar h}_j\}$ and $\{{\bar h}'_j\}$ are continuous everywhere, in particular at the similarity horizon. \\
(3) $\{{\bar h}''_j\}$ and $\{\dot{\bar h}''_j\}$ are discontinuous at the similarity horizon, although they have finite one-sided limit values as $X\to X_{\rm (s)}-0$ and $X\to X_{\rm (s)}+0$.\\
We denote the full-order perturbations as
\begin{eqnarray}
\delta g(U,X)&\equiv& g(U,X)-g_{\rm (b)}(X),\\
\delta {\bar g}(U,X)&\equiv& {\bar g}(U,X)-{\bar g}_{\rm (b)}(X), \\
\delta {\bar h}_j(U,X)&\equiv& {\bar h}_j(U,X)-{\bar h}_{j\rm (b)}(X),
\end{eqnarray}
where $g_{\rm (b)}$, ${\bar g}_{\rm (b)}$ and $\{{\bar h}_{j\rm (b)}\}$ denote the background self-similar solution.
 
Now we consider the behavior of the perturbations at the similarity
horizon, of which we mean $X\to X_{\rm (s)}-0$ for $g_{\rm (s)}{\bar
g}_{\rm (s)}>0$, while $X\to X_{\rm (s)}+0$ for $g_{\rm (s)}{\bar
g}_{\rm (s)}<0$. 
Applying the similar analysis as that in~\cite{hm2003}, we finally obtain the full-order perturbation equation for $\delta {\bar h}_{j}^{''}$ at the similarity horizon is obtained as
\begin{equation}
\delta\dot{{\bar h}}_{j}^{''}=\left(-1+8\pi\sum^n_{i=1}\kappa_{i}^2\right)\delta {\bar h}_{j}^{''}.
\end{equation}
This equation can be integrated to obtain
\begin{eqnarray}
\delta {\bar h}_{j}^{''} \propto e^{\beta U},
\end{eqnarray}
where
\begin{eqnarray}
\beta \equiv -1+8\pi\sum^n_{i=1}\kappa_{i}^2.
\end{eqnarray}
Therefore, for early-time solutions, it is found that the perturbation decays exponentially for $8\pi\sum^n_{i=1}\kappa_{i}^2 <1$, it is constant for $8\pi\sum^n_{i=1}\kappa_{i}^2=1$ and it grows exponentially for $8\pi\sum^n_{i=1}\kappa_{i}^2 > 1$. The situation is reversed for late-time solutions.

It is noted that these perturbations are gauge-independent as shown in Appendix~\ref{gauge}. Here we define instability by the exponential growth of discontinuity. Then we find the following criterion: for early-time solutions, the solutions with regular similarity horizon and $8\pi\sum^n_{i=1}\kappa_{i}^2<1$ are stable against the kink mode, while those with $8\pi\sum^n_{i=1}\kappa_{i}^2 >1$ are unstable. Solutions with $8\pi\sum^n_{i=1}\kappa_{i}^2=1$ are marginally stable against this mode. The situation is reversed for late-time solutions. Setting $n=1$, we reproduce the result obtained in~\cite{hm2003} for the case of a massless scalar field.

\section{power-law flat FRW spacetime as a self-similar spacetime}
\label{ssFRW}
We prove that if the flat FRW spacetime is self-similar, 
then the scale factor follows a power-law.
We show that the flat FRW spacetime with a metric
\begin{eqnarray}
ds^2=-dt^{2}+a(t)^2(dr^2+r^{2}d\Omega^2), \label{FFRW}
\end{eqnarray}
can be transformed to a general spherically symmetric metric
\begin{equation}
ds^2=-A(z)^2dt^2+B(z)^2d{\bar r}^2+C(z)^2 {\bar r}^2 d\Omega^2,\label{ssform}
\end{equation}
with a homothetic Killing vector
\begin{eqnarray}
t \frac{\partial}{\partial t}+{\bar r} \frac{\partial}{\partial {\bar r}},
\end{eqnarray}
where $z \equiv t/{\bar r}$, if and only if the scale 
factor $a(t)$ obeys the power-law. We consider coordinate transformations from Eq.~(\ref{FFRW}) to a self-similar form (\ref{ssform}). $t$ remains unchanged, apparently, while we define a new coordinate ${\bar r}={\bar r}(r)$ to obey
\begin{eqnarray}
a(t)^2 dr^2 = B(z)^2 d{\bar r}^2,
\end{eqnarray}
and
\begin{eqnarray}
a(t)^2 r^2  = C(z)^2 {\bar r}^2.
\end{eqnarray}
Using a relation $t=z{\bar r}$ and setting $z=1$, we can rewrite these equations as 
\begin{eqnarray}
a^2 ({\bar r})&=& B(1)^2 (d{\bar r}/dr)^2,
\end{eqnarray}
and 
\begin{eqnarray}
a^2 ({\bar r})&=& C(1)^2 ({\bar r}/r)^2. \label{ap1}
\end{eqnarray}
These equations result in 
\begin{eqnarray}
\left(\frac{d{\bar r}}{dr}\right)^2=\left(\frac{C(1)}{B(1)}\right)^2\left(\frac{{\bar r}}{r}\right)^2.
\end{eqnarray}
This equation can be integrated to obtain
\begin{eqnarray}
{\bar r} \propto r^s,
\end{eqnarray}
where $s\equiv C(1)/B(1)$.
Then Eq.~(\ref{ap1}) gives
\begin{eqnarray}
a^2 ({\bar r}) \propto {\bar r}^{2(1-1/s)},
\end{eqnarray}
and therefore we conclude that the scale factor obeys the power-law:
\begin{eqnarray}
a(t) \propto t^{(1-1/s)}.
\end{eqnarray}

Inversely, as shown in \S~\ref{FRWbasic}, we prove that if the flat FRW spacetime has a scale
factor which follows the power-law, then the spacetime is self-similar with $\alpha \ne 1$.

It is noted that the case with $\alpha=1$ is also homothetic, where the homothetic vector 
is given by $t\partial/\partial t $~\cite{mh2004}.

\section{gauge-invariance of kink instability}
\label{gauge}
To construct the gauge-invariant quantity we introduce the $2+2$ split of spherically symmetric spacetimes according to~\cite{gs}. We write the spherically symmetric spacetime as a product manifold ${\cal M}=M^2\times S^2$ with metric
\begin{equation}
g_{\mu\nu}=\mbox{diag}(g_{AB},r^2\gamma_{ab}),
\end{equation}
where $g_{AB}$ is an arbitrary Lorentz metric on $M^2$, $r$ is a scalar on $M^2$ with $r=0$ defining the boundary of $M^2$, and $\gamma_{ab}$ is the unit curvature metric on $S^2$. We introduce the covariant derivatives on spacetime ${\cal M}$, the subspacetime $M^2$ and the unit sphere $S^2$ with
\begin{eqnarray}
g_{\mu\nu;\lambda}&=&0,\\
g_{AB|C}&=&0,\\
g_{ab:c}&=&0.
\end{eqnarray}
We also introduce totally antisymmetric covariant tensors on ${\cal M}$, $M^2$ and $S^2$, respectively:
\begin{eqnarray}
&&\varepsilon_{\mu\nu}=-\varepsilon_{\nu\mu},~~\varepsilon_{\mu\nu;\lambda}=0,~~\varepsilon_{\mu\lambda}\varepsilon^{\nu\lambda}=-g_\mu^\nu, \\
&&\varepsilon_{AB}=-\varepsilon_{BA},~~\varepsilon_{AB|C}=0,~~\varepsilon_{AC}\varepsilon^{BC}=-g_A^B, \\
&&\varepsilon_{ab}=-\varepsilon_{ba},~~\varepsilon_{ab:c}=0,~~\varepsilon_{ac}\varepsilon^{bc}=-g_a^b.
\end{eqnarray}

The kink mode concerns the density gradient on the comoving slicing for the fluid case, which is gauge-invariant~\cite{harada2001,hm2003}. For the scalar field case, it concerns the second-order derivative of the scalar field~\cite{hm2003}. From the second-order derivative, we can construct a similar quantity $I$ such as
\begin{equation}
I=\varepsilon^{AB}g^{CD}\phi_{j|B}\phi_{j|D}\phi_{j|AC},
\end{equation}
for some $j$. Since the metric, the scalar field and their first derivatives with respect to $T$ and $X$ are not perturbed at the similarity horizon for the kink perturbation, the full-order perturbation $\delta I \equiv I-I_{\rm (b)}$ becomes
\begin{equation}
\delta I=\varepsilon^{AB}g^{CD}\phi_{j|B}\phi_{j|D}\delta \phi_{j|AC},
\end{equation}
where $\delta \phi_j =\phi_j-\phi_{j\rm (b)}$. 

For the perturbed FRW universe given in \S~\ref{body}, the line element on $M^2$ is
\begin{eqnarray}
ds_{M^2}^2&=&-(e^{2\sigma(T,Z)}-e^{2(\alpha-1)Z+2\omega(T,Z)})e^{2T}dT^2 \nonumber \\
&&-2e^{2(\alpha-1)Z+2T+2\omega}dTdZ \nonumber\\
&&+e^{2(\alpha-1)Z+2T+2\omega}dZ^2.
\end{eqnarray}
Therefore, it is clear that for the background flat FRW universe, the quantity $I$ has a nontrivial value $I_{\rm (b)}$ even at the similarity horizon, i.e. 
\begin{eqnarray}
I_{\rm (b)}=-2\alpha\kappa_j^3e^{-2T}.
\end{eqnarray}
$\delta I$ in the limit to the similarity horizon is calculated to be 
\begin{equation}
\delta I=\kappa_j^2e^{-2T}\psi''_{j\rm (s)}.
\end{equation}
The factor $e^{-2T}$ is the same as that for the background flat FRW universe. Therefore, we obtain for the perturbation evolution in the limit to the similarity horizon 
\begin{equation}
\frac{\delta I}{I_b} \propto \psi''_{j\rm (s)}(T).
\end{equation}

For more general case in Appendix~\ref{general}, the line element on $M^2$  can be written as 
\begin{eqnarray}
ds_{M^2}^2=gxe^{-2U}\biggl[\biggl(2-\frac{{\bar g}}{x}\biggl)dU^2+2dUdX\biggl].
\end{eqnarray}
Therefore, the quantity $I$ has a nontrivial value $I_{\rm (b)}$ at the similarity horizon, i.e. 
\begin{eqnarray}
I_{\rm (b)}=e^{2U}f(X),
\end{eqnarray}
where $f(X)$ is a function of $X$.

In terms of the coordinates $(U,X)$ presented in Appendix~\ref{general}, $\delta I$ in the limit to the similarity horizon is calculated to be 
\begin{equation}
\delta I=-\frac{e^{2U}}{g_{\rm (s)}x_{\rm (s)}}{\dot \phi}_{j\rm (s)}^2\delta \phi''_{j\rm (s)}.
\end{equation}
Or, using the notation presented in Appendix~\ref{general}, we obtain
\begin{equation}
\delta I=-\frac{e^{2U}}{g_{\rm (s)}x_{\rm (s)}}\kappa_j^2\delta 
\phi''_{j\rm (s)},
\end{equation}
in the limit to the similarity horizon. The factor $e^{2U}$ is the same as that for the background solution. Therefore, we obtain for the perturbation evolution in the limit to the similarity horizon 
\begin{equation}
\frac{\delta I}{I_b} \propto \delta \phi''_{j\rm (s)}(U).
\end{equation}
 
In the whole spacetime ${\cal M}$ the quantity $I$ can be written as 
\begin{equation}
I=\varepsilon^{\mu\nu}g^{\lambda\sigma}\phi_{j;\nu}\phi_{j;\sigma}\phi_{j;\mu\lambda}.
\end{equation}
It is clear that this is gauge-independent.

\newpage
\begin{table}[htbp]
\begin{center}
\caption{\label{tb:perfect_fluid}
Stability of kink mode perturbation in 
the flat FRW universe with a perfect fluid
with the equation of state $p=k\mu$.
The power index $\alpha$ of the scale factor is given by
$\alpha=2/[3(1+k)]$.
See~\cite{harada2001} for details.}
\begin{tabular}{|c|c|c|c|}
\hline 
Parameter & Power index $\alpha$ & Stability \\
\hline
$0<k<1/3$   & $1/2<\alpha<2/3 $ & Stable \\
$k=1/3$ & $\alpha =1/2$ & Unstable\footnote{In this case, although 
the linear stability is marginal, nonlinear effect causes 
nonexponential instability.} \\
$1/3<k\le 1$ & $1/3\le \alpha <1/2$ & Unstable \\
\hline 
\end{tabular}
\end{center}
\end{table}

\begin{table}[htbp]
\begin{center}
\caption{\label{tb:scalar_field}
Stability of kink mode perturbation in 
the flat FRW universe 
with scalar fields. For exponential potentials,
we consider the sum of independent potentials,
$V_{\rm tot}=\sum_{i}V_{i}\exp(-\sqrt{8\pi}\lambda_{i}\phi_{i})$,
for scalar fields $\{\phi_{i}\}$, where
the power index $\alpha$ of the scale factor is given by
$\alpha=2\sum_{i}\lambda_{i}^{-2}$.}
\begin{tabular}{|c|c|c|c|}
\hline 
Potential & Parameters & Power index $\alpha$ & Stability \\
\hline 
Massless & --- & $\alpha=1/3$ & Unstable \\
\hline 
     & $\sum_{i}\lambda_{i}^{-2}>1/4$ & $\alpha>1/2$ & Stable \\
Exp & $\sum_{i}\lambda_{i}^{-2}=1/4$ & $\alpha=1/2$ & Marginal\\
     & $1/6<\sum_{i}\lambda_{i}^{-2}<1/4$ 
& $1/3<\alpha<1/2$ & Unstable \\ 
\hline 
\end{tabular}
\end{center}
\end{table}

\end{document}